\documentclass[a4paper,fleqn]{cas-dc}

\usepackage[numbers]{natbib}
\usepackage{tabularx}
\usepackage{booktabs}
\usepackage{makecell}
\usepackage{caption}
\usepackage{microtype}
\usepackage{multirow}
\def\tsc#1{\csdef{#1}{\textsc{\lowercase{#1}}\xspace}}
\tsc{WGM}
\tsc{QE}
\tsc{EP}
\tsc{PMS}
\tsc{BEC}
\tsc{DE}

\begin{document}
\let\WriteBookmarks\relax
\def\floatpagepagefraction{1}
\def\textpagefraction{.001}
\shorttitle{SCA}
\shortauthors{Asi Khandelwal  et al.}

\title [mode = title] {Improving the critical current density of the \texorpdfstring{$\mathrm{V}_{0.59}\mathrm{Ti}_{0.40}\mathrm{Ce}_{0.01}$}{V0.59Ti0.40Ce0.01} alloy superconductor through successive cold-working and annealing at different temperatures}

\author[1,2]{Asi Khandelwal}
[% 
orcid=0000-0002-7756-5574]
\credit{Investigation, Data curation, Formal analysis, Writing - Original draft preparation}
\address[1]{Free Electron Laser Utilization Laboratory, Raja Ramanna Centre for Advanced Technology, Indore - 452013, India}
\address[2]{Homi Bhabha National Institute, Training School Complex, Anushakti Nagar, Mumbai - 400094, India}
\ead{asikhandelwal1503@gmail.com}

\author[3]{SK. Ramjan}
\address[3] {European Organization for Nuclear Research (CERN), CH-1211 Geneva, Switzerland}
\credit{Investigation}

\author[1,4]{Basudev Padhi}
\address[4] {School of Studies in Physics, Vikram University, Ujjain-456010, India}
\credit{Investigation}

\author[1,2]{L. S. Sharath Chandra}[%
orcid=0000-0002-1253-6035]
\cormark[1]
\ead{lsschandra@rrcat.gov.in}
\credit{Conceptualization of this study, Formal analysis, Writing - Review & Editing}

\author[2,5]{Archna Sagdeo}%
\address[5] {Accelerator Physics and Synchrotrons Utilization Division, Raja Ramanna Centre for Advanced Technology, Indore-452 013, India}
\credit{Investigation}

\author[6]{Kranti Kumar}%
\address[6] {UGC-DAE Consortium for Scientific Research, Khandwa Road, Indore 452009, India}
\credit{Investigation}

\author[6]{Sudip Pal}%

\author[1,2] {M. K. Chattopadhyay}
 \credit{Writing - Review & Editing, Project administration}

\begin{abstract}
The critical current density ($J_c$) of V--Ti alloy superconductors is strongly influenced by the size and distribution of microstructural defects that effectively pin magnetic flux lines. In this work, the effect of successive cold working and annealing (SCA) at 550~$^{\circ}$C on the microstructure, superconducting properties, and flux-pinning behaviour of the V$_{0.59}$Ti$_{0.40}$Ce$_{0.01}$ alloy is investigated and compared with the previously reported SCA at 450 and 650~$^{\circ}$C. During the SCA at 550~$^{\circ}$C, the superconducting transition temperature ($T_c$) increases gradually with successive processing steps. The first annealing after cold rolling to 50\% thickness produces a significant enhancement in $J_c$ over the entire measured magnetic field range, whereas the subsequent SCA cycles result in only marginal changes. The $J_c$ remains relatively weakly dependent on magnetic field over a wide magnetic-field range, and the final cold-worked sample exhibits a finite $J_c$ up to 8.5 T. Pinning force density analysis shows that grain boundaries dominate the flux pinning in low magnetic fields, whereas dislocations and $\beta$--$\alpha'$ interfaces become the dominant pinning centres in higher fields. Comparison of the different SCA temperatures shows that 650~$^{\circ}$C provides the highest low-field $J_c$, whereas 450~$^{\circ}$C gives the best high-field performance. In contrast, SCA at 550~$^{\circ}$C provides the most balanced field dependence and the largest enhancement in $J_c$ relative to the corresponding as-cast alloy. Although the highest absolute $J_c$ in the high-field regime is achieved after the third SCA cycle at 450~$^{\circ}$C, a comparable value is obtained after only the first annealing at 550~$^{\circ}$C. These results demonstrate that the intermediate annealing temperature of 550~$^{\circ}$C provides an effective balance between defect generation, phase evolution, and recovery, resulting in enhanced flux pinning over a wide magnetic-field range.
\end{abstract}

%\date{}

\maketitle

\section{INTRODUCTION}

Superconducting magnets are indispensable for generating high magnetic fields with negligible electrical losses. They are widely used in magnetic resonance imaging, particle accelerators, fusion reactors, nuclear magnetic resonance spectroscopy, transportation, and scientific research \cite{strickland, ruiz, garcia, kingham,  prikhna, kushwaha, Banno2023}. The performance of these magnets is primarily determined by the critical current density ($J_c$), which defines the maximum dissipation-free current a superconductor can carry \cite{ruiz}. Although NbTi and Nb$_3$Sn remain the dominant commercial superconductors, their use in neutron-irradiation environments is limited because irradiation degrades their superconducting properties and produces long-lived radioactive isotopes \cite{Gajda2025, Nishimura}. 
More recently, MgB$_2$, rare-earth barium copper oxide (REBCO) coated conductors, iron-based superconductors and high-entropy alloy superconductors have emerged as promising alternatives because of their higher operating temperatures and excellent superconducting performance \cite{prikhna, kitagawa, godeke, sharma, Baumann2026, jangid, aj, guo,sene, idc, Lei2023}. Despite these advances, the search for alternative superconducting materials and effective processing strategies for existing superconductors remains an active area of research \cite{ ruiz, Banno2023, prikhna, aj, guo,sene, Lei2023}. Among metallic low-temperature superconductors, the $\beta$-V--Ti alloys have emerged as promising materials for fusion and other radiation-intensive applications owing to their excellent radiation resistance, short-lived activation products under neutron irradiation, and good mechanical workability \cite{Nag, butt, Zhang, tak10, tai}. However, their relatively low $J_c$ compared with commercial Nb-based superconductors remains the primary obstacle to their practical applications \cite{ mat15, mat13,vzr}.

Improving the $J_c$ in superconductors largely relies on enhancing the vortex pinning through microstructural engineering \cite{speller, ji}. Tailoring the type, size, density, and distribution of defects by alloying, thermomechanical processing and heat treatment can introduce effective flux-pinning centres and substantially improve the superconducting performance \cite{ji}. This approach has been successfully demonstrated in Nb--Ti, Nb$_3$Sn, MgB$_2$, REBCO coated conductors and high-entropy alloy superconductors \cite{Gajda2025, kitagawa, yan, ersoz, kumar,LPBf, hidayati, kim, shadab, simon2024, geng}. Among these strategies, thermomechanical processing through repeated deformation and heat treatment cycles is particularly attractive because it modifies the defect structure without altering the alloy chemistry and closely mimics the industrial wire fabrication process \cite{Lei2023, speller, ji, yz, zhu}.

In the $\beta$-V--Ti alloys, grain boundaries dominate flux pinning in low magnetic fields, whereas dislocations become increasingly effective in higher fields \cite{mat13, mat15}. More recently, the addition of rare-earth elements with negligible solubility in the V--Ti matrix has been shown to further enhance the $J_c$ by promoting grain refinement and the formation of rare-earth-rich precipitates that provide additional pinning centres \cite{gd, y, si, sri}. In particular, Ce and Gd additions have been found to be the most effective in enhancing vortex pinning \cite{sca-450, sca-650}.

Earlier, we have shown that successive cold working and annealing (SCA) at 450~$^\circ$C significantly improves the $J_c$ of the V$_{0.59}$Ti$_{0.40}$Ce$_{0.01}$ alloy, even in high magnetic fields \cite{sca-450}. In contrast, SCA at 650~$^{\circ}$C yields the highest low-field $J_c$, while the high-field $J_c$ deteriorates rapidly \cite{sca-650}. This deterioration is attributed to the growth of the $\alpha$-phase precipitates beyond the vortex-core size and the reduction in dislocation density due to recrystallization \cite{sca-650}. These results demonstrate the strong influence of annealing temperature on the evolution of the pinning microstructure and the resulting superconducting performance. However, the achieved $J_c$ values remain lower than those of commercial superconductors. Therefore, identifying an optimum annealing temperature that simultaneously maximizes both low- and high-field $J_c$ remains an important challenge. 

To address this, the present work investigates SCA at 550~$^\circ$C with an annealing duration of 5 hours, consistent with the previously investigated SCA series. As an intermediate annealing temperature, 550~$^\circ$C is expected to promote the formation of the $\alpha$ phase, which is absent after annealing at 450~$^\circ$C, while suppressing its excessive growth observed at 650~$^\circ$C. It is also expected to retain a higher dislocation density than annealing at 650~$^\circ$C, thereby preserving effective high-field pinning. The resulting microstructure, electrical resistivity, and $J_c$ were systematically investigated and compared with those reported previously after SCA at 450~$^\circ$C and 650~$^\circ$C.

\section{EXPERIMENTAL DETAILS}

\begin{figure}
\includegraphics[width= 1\columnwidth]{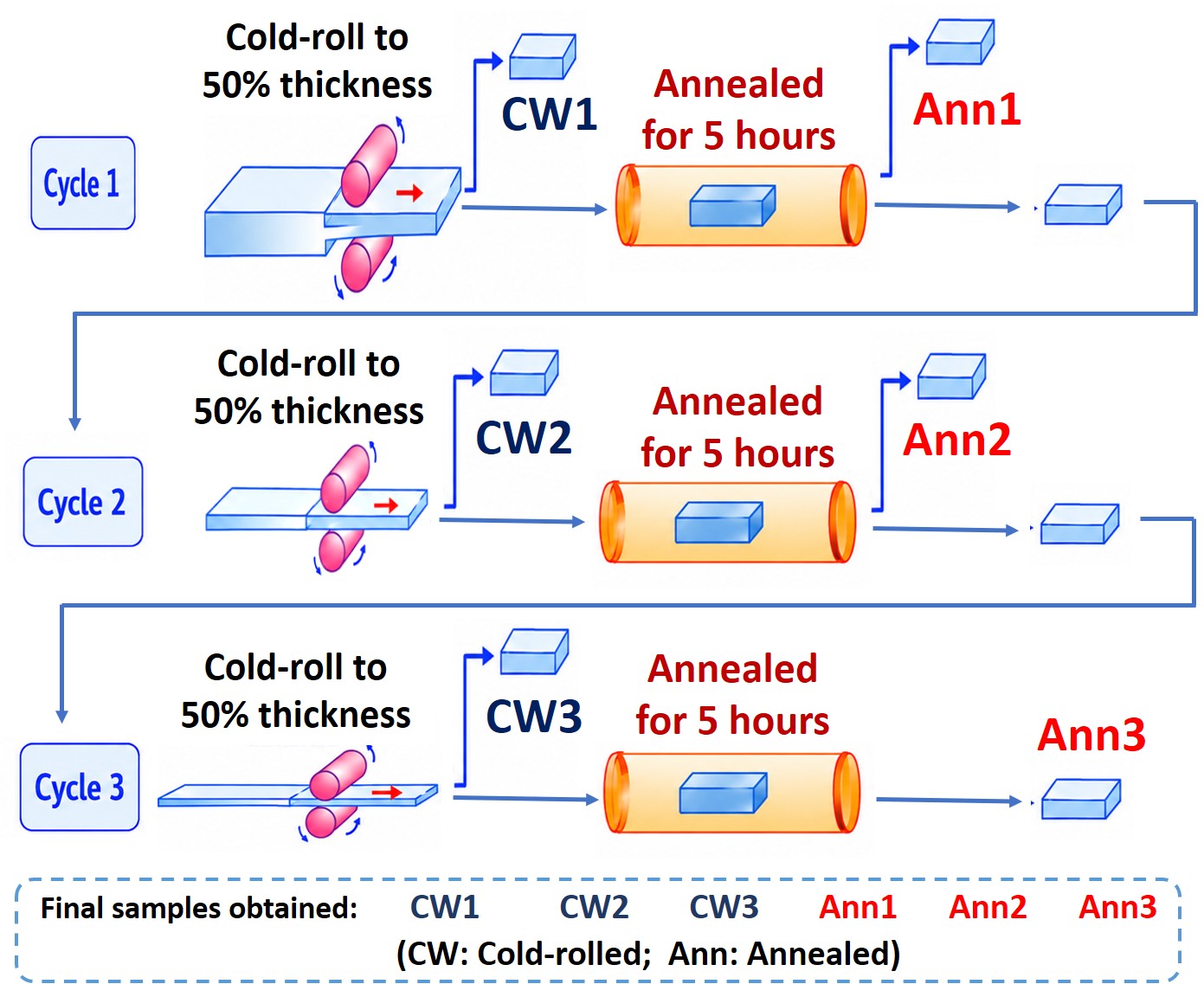}

\caption{ Schematic illustration of the successive cold-working and annealing process adopted for the V$_{0.59}$Ti$_{0.40}$Ce$_{0.01}$ alloy.}

\label{sample prep}
\end{figure}

The polycrystalline V$_{0.59}$Ti$_{0.40}$Ce$_{0.01}$ alloy superconductors were prepared by arc-melting \cite{sca-650}. A diamond saw was used to cut the samples, and the resulting disks were subjected to SCA \cite{sca-650}. Each SCA cycle consisted of cold rolling to 50\% of its prior thickness, followed by annealing at 550~$^{\circ}$C for 5 hours and slow cooling down to room temperature. Portions of the sample were retained after each cold-working and annealing step, yielding the CW1, Ann1, CW2, Ann2, CW3, and Ann3 samples. The complete SCA route is schematically illustrated in Fig.~\ref{sample prep}.

The X-ray diffraction (XRD) measurements were performed using radiation with a wavelength of 0.725 Å from the BL-12 beamline of the Indus-2 synchrotron facility. The electrical resistivity measurements were carried out using a cryogen-free magnet cryostat (CFM, Cryogenics, UK). The magnetization (M) measurements were performed using a Superconducting Quantum Interference Device-based Vibrating Sample Magnetometer (MPMS-3 SQUID VSM, Quantum Design, USA) and a 16 T vibrating sample magnetometer (VSM, Quantum Design, USA).

\section{RESULTS AND DISCUSSION}

SCA increases the grain-boundary (GB) and dislocation densities, primarily during the cold-working steps. Annealing induces recrystallization and phase transformations, and the evolution of these microstructural features depends strongly on the annealing temperature. These defects act as effective flux line pinning centres and therefore strongly influence the superconducting properties. The results obtained after SCA at 550~$^\circ$C are discussed first, and subsequently compared with those reported earlier for SCA at 450 and 650~$^\circ$C.

\subsection{Effect of Successive Cold Working and Annealing at 550~$^{\circ}$C}

\begin{figure}
\includegraphics[width= 1\columnwidth]{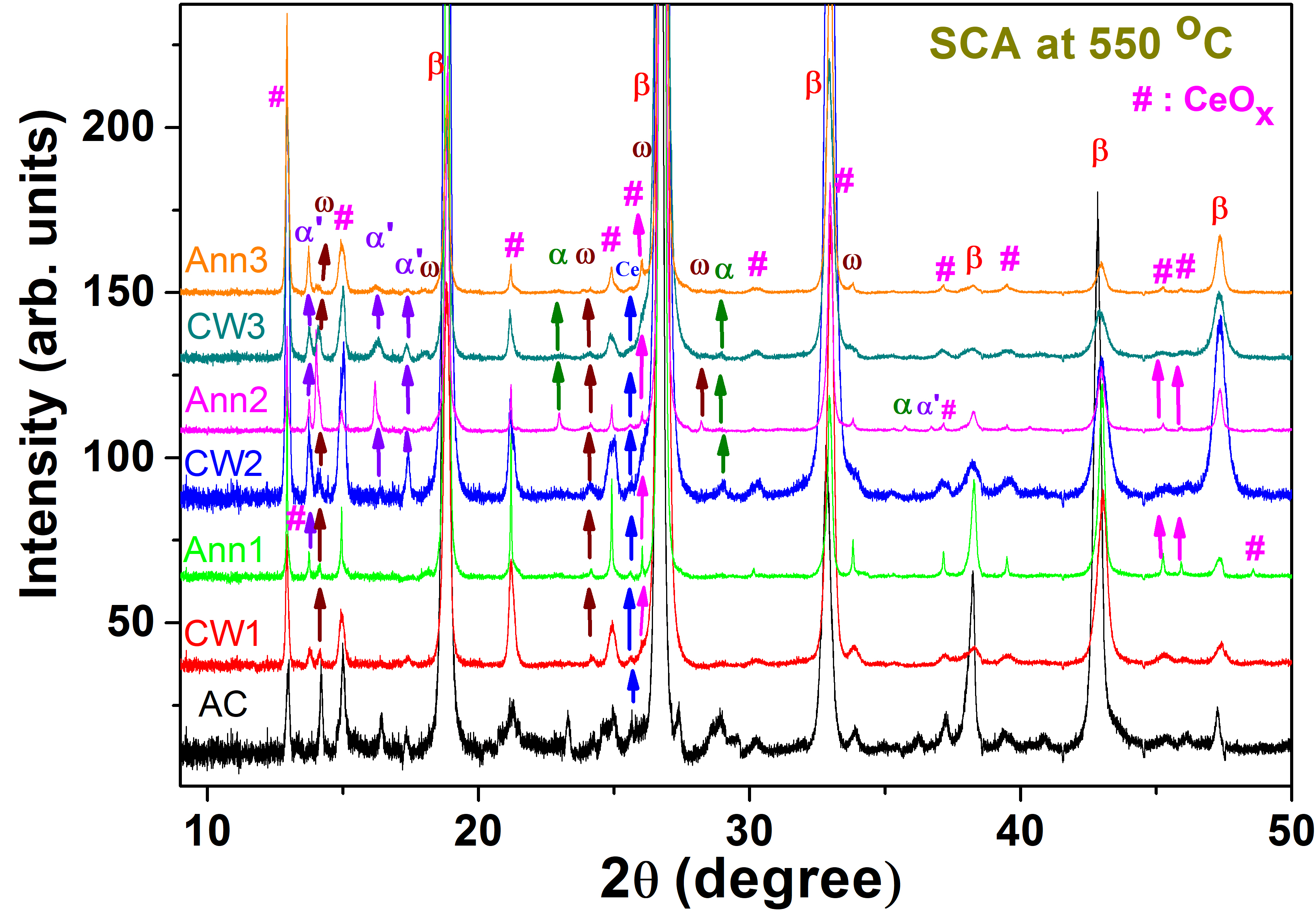}

\caption{ XRD patterns of the V$_{0.59}$Ti$_{0.40}$Ce$_{0.01}$ alloy at different stages of SCA at 550$^{\circ}$C. The diffraction peaks are indexed to the $\beta$, $\omega$, $\alpha$, and $\alpha'$-phases, cerium oxide, and elemental Ce.}

\label{xrd}
\end{figure}

Figure \ref{xrd} shows the room-temperature XRD patterns of the V$_{0.59}$Ti$_{0.40}$Ce$_{0.01}$ alloy at different stages of SCA at 550~$^{\circ}$C. All the samples predominantly consist of the body-centered cubic $\beta$ phase, accompanied by the hexagonal close-packed $\omega$ and $\alpha$ phases, the orthorhombic $\alpha'$ phase, and the face-centered cubic (fcc) cerium oxide. In addition, a weak diffraction peak corresponding to elemental Ce (fcc) is also observed. The diffraction patterns were analysed using the PowderCell software \cite{POWDER} to index the major peaks and estimate the lattice parameters of all phases. The lattice parameter of the $\beta$ phase decreases slightly from 3.14 to 3.13~\AA\ after the first cold-working step and remains nearly unchanged thereafter. Likewise, the lattice parameters of the secondary phases show only minor variations throughout the SCA process, indicating that their crystal structures remain essentially unchanged. Similarly, the lattice parameter of the cerium oxide phase remains in the range of 5.56--5.58~\AA, which is larger than that of stoichiometric CeO$_2$ (5.4~\AA) \cite{chiang}. Since oxygen non-stoichiometry increases the lattice parameter of cerium oxide \cite{chiang,ceo}, the phase is identified as CeO$_x$ \cite{sca-650}.

\begin{figure}
\includegraphics[width= 1\columnwidth]{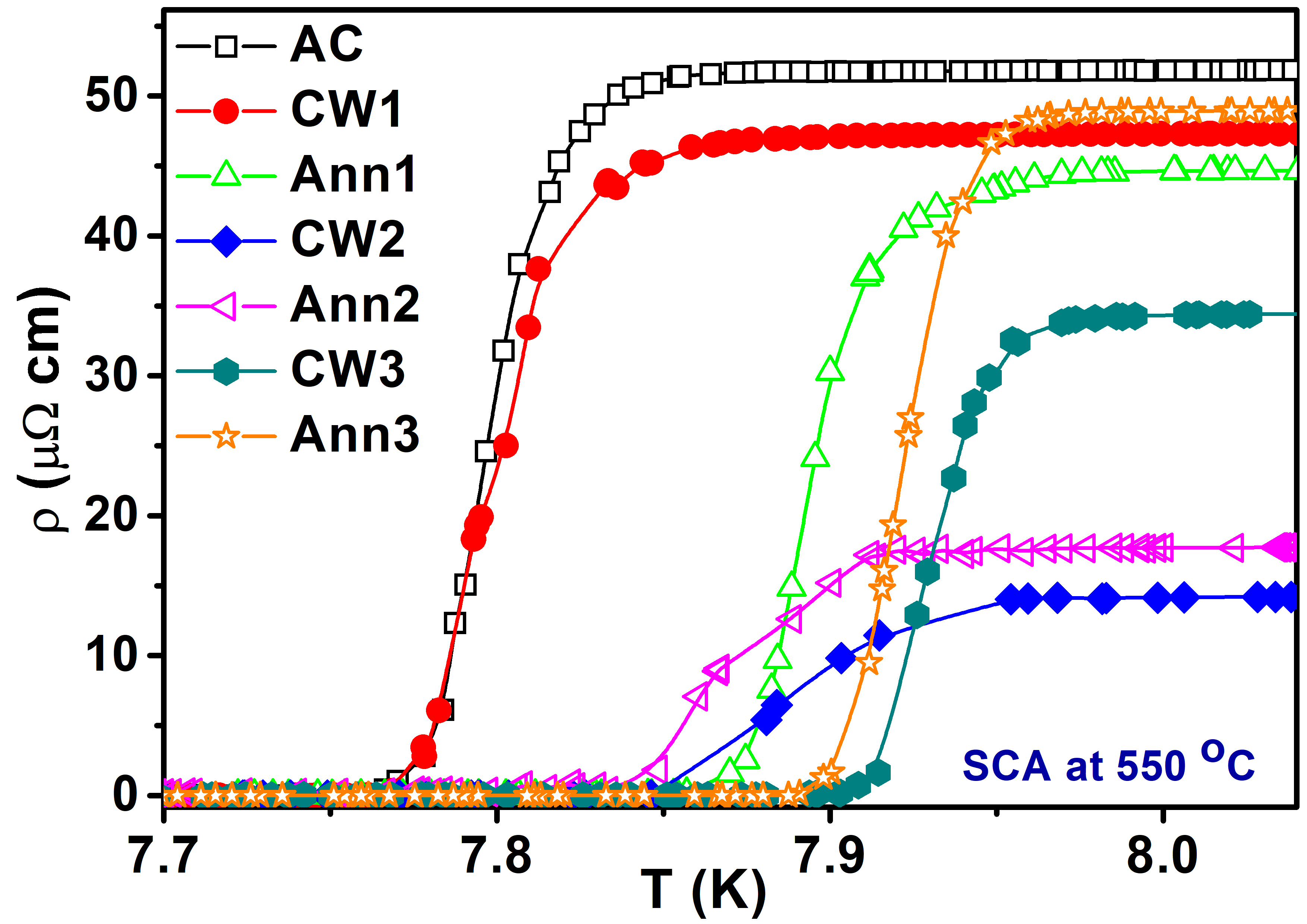}

\caption{ Temperature dependence of electrical resistivity of the V$_{0.59}$Ti$_{0.40}$Ce$_{0.01}$ alloy at different stages of SCA at 550$^{\circ}$C. The \textit{T$_c$} increases with SCA.}
0
\label{tc}

\end{figure}

Figure \ref{tc} shows the temperature dependence of electrical resistivity ($\rho$($T$)) for the V$_{0.59}$Ti$_{0.40}$Ce$_{0.01}$ alloy at different stages of SCA at 550~$^{\circ}$C. The superconducting transition temperature (\textit{T$_c$}) increases with SCA, probably due to the increase of volume fraction of the superconducting $\alpha'$-phase \cite{sca-650}. The \textit{T$_c$} is defined as the temperature at which the temperature derivative of resistivity shows a maximum. During SCA at 550~$^{\circ}$C, the residual resistivity ($\rho_0$) decreases initially, reaches a minimum at CW2, and then increases during the subsequent processing steps.

\begin{figure}
\includegraphics[width= 1\columnwidth]{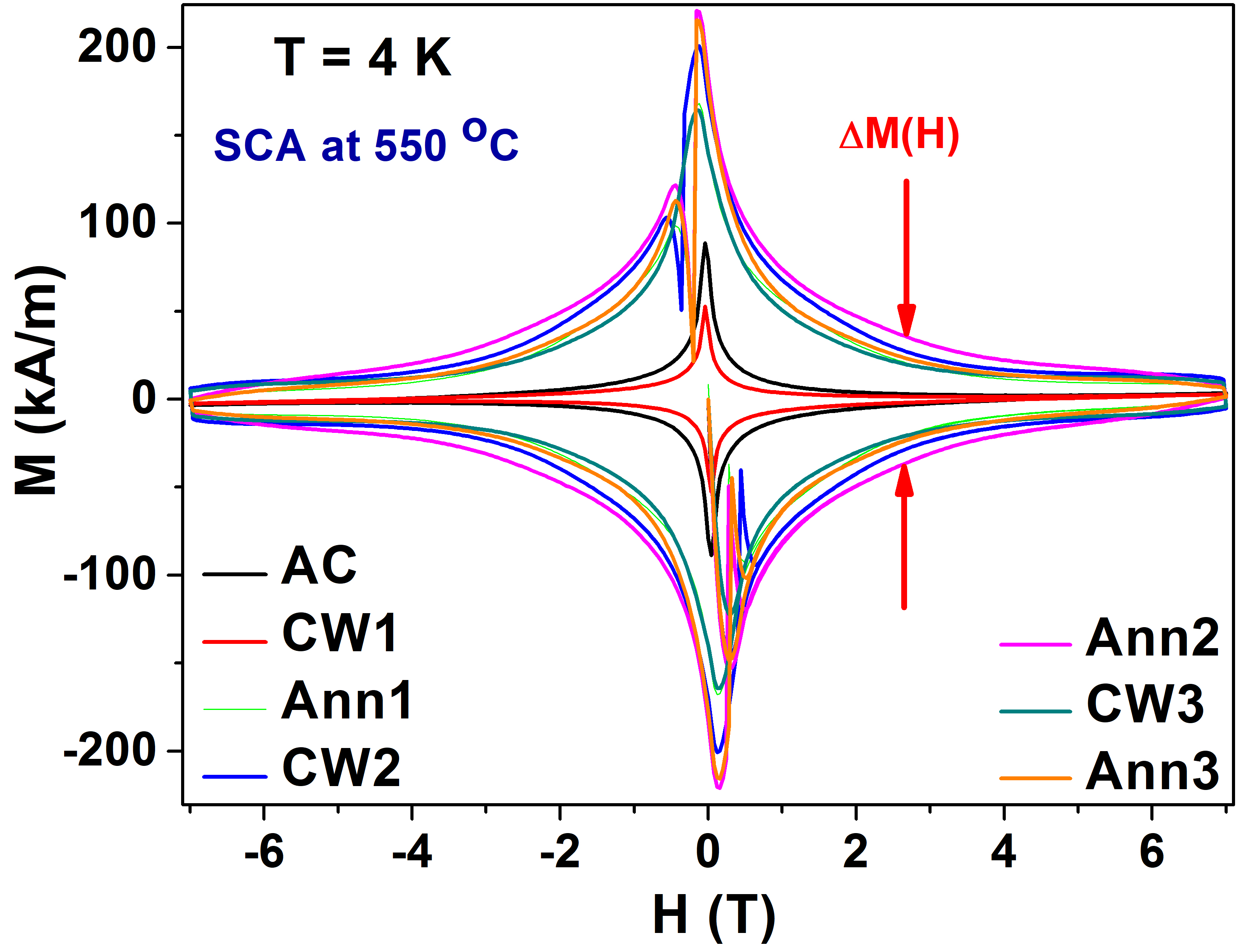}

\caption{ Isothermal field dependence of the magnetization of the V$_{0.59}$Ti$_{0.40}$Ce$_{0.01}$ alloys at different stages of SCA measured at 4 K.}

\label{mh}

\end{figure}

\begin{figure}
\includegraphics[width= 1\columnwidth]{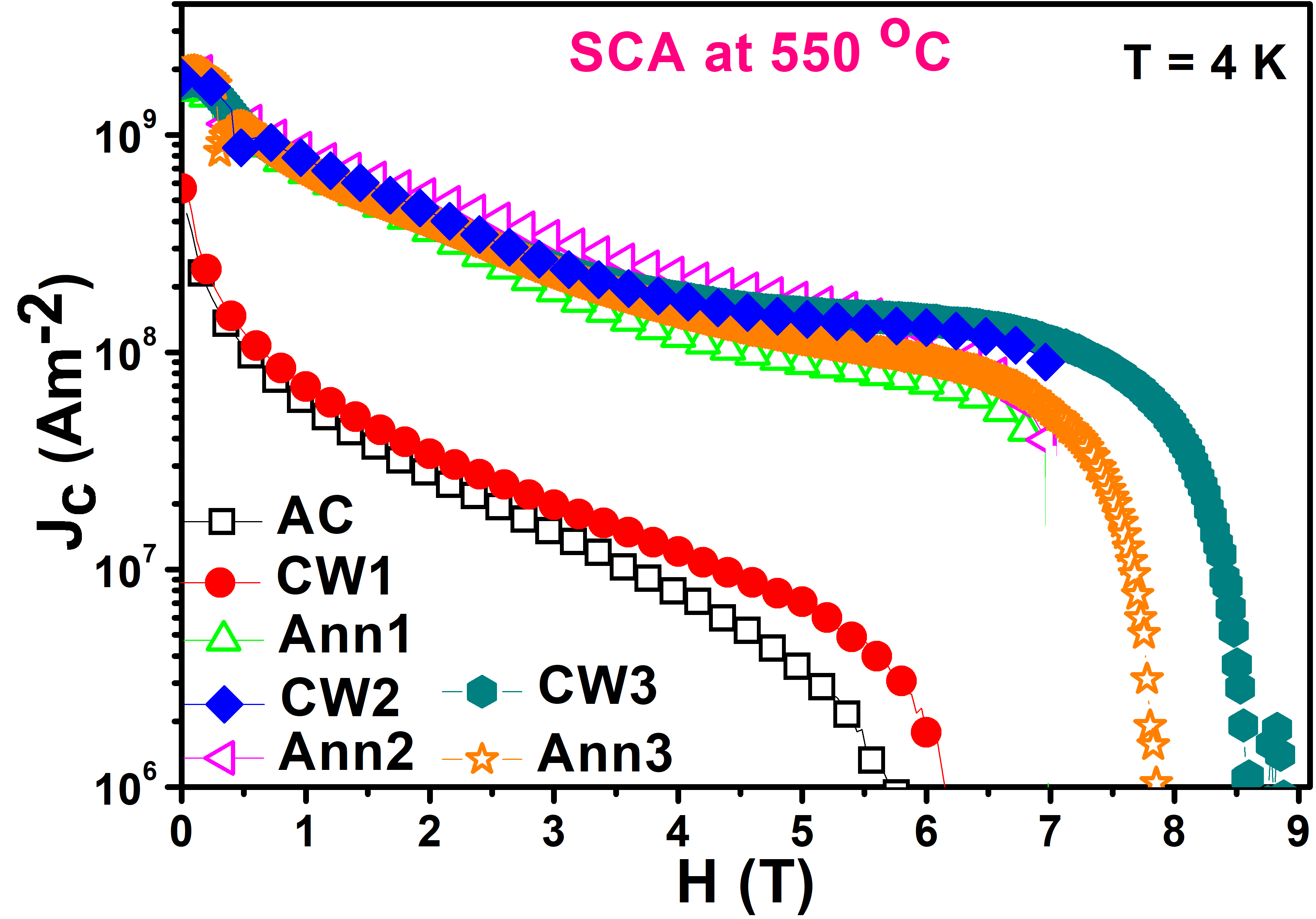}

\caption{Isothermal field dependence of the $J_c$ of the V$_{0.59}$Ti$_{0.40}$Ce$_{0.01}$ alloy at different stages of SCA, measured at 4 K. After Ann1, $J_c$ increases significantly over the entire measured field range, whereas the subsequent SCA cycles result in only marginal changes. The CW3 sample exhibits a finite $J_c$ up to 8.5 T.}

\label{jc}

\end{figure}

The field dependence of magnetization ($M(H)$) measured at 4 K for the V$_{0.59}$Ti$_{0.40}$Ce$_{0.01}$ alloy at different stages of SCA at 550~$^{\circ}$C is shown in Fig.~\ref{mh}. All samples exhibit magnetic hysteresis due to irreversible flux penetration and vortex pinning. At a given magnetic field, the difference in magnetization measured during the field-decreasing and subsequent field-increasing cycles ($\Delta M$) was used to calculate $J_c$ using the Bean critical-state model \cite{sca-650,bean64}. For a rectangular sample, the critical current density is given by $J_c = 2\Delta M[t(1-t/3w)]^{-1}$,  where $t$ and $w$ ($w>t$) denote the dimensions of the sample perpendicular to the applied magnetic field. Figure \ref{jc} shows the isothermal field dependence of $J_c$ at 4 K of the V$_{0.59}$Ti$_{0.40}$Ce$_{0.01}$ alloy at different stages of SCA at 550~$^{\circ}$C. The maximum $J_c$ for the as-cast alloy is 4.35 $\times$ 10$^{8}$~A/m$^{2}$ in zero magnetic field and 1.6 $\times$ 10$^{4}$~A/m$^{2}$ in 7 T. After CW1, the $J_c$ remains nearly the same as that of the as-cast sample. However, after Ann1, the $J_c$ increases significantly over the entire field range of measurement, reaching 1.61 $\times$ 10$^{9}$~A/m$^{2}$ in zero field and 3.38 $\times$ 10$^{7}$~A/m$^{2}$ in 7 T. The subsequent SCA steps produce only marginal changes in $J_c$ across the entire field range. Furthermore, for the samples processed by SCA at 550~$^{\circ}$C, the $J_c$ remains significant even in 7 T field and does not decrease rapidly with increasing magnetic field. Therefore, the magnetization measurements for the CW3 and Ann3 samples were extended to higher magnetic fields. The Ann3 sample shows a measurable $J_c$ up to about 8 T, whereas the CW3 sample exhibits a finite $J_c$ up to 8.5 T. The superior high-field performance of CW3 is attributed to the higher dislocation density retained after cold working, which provides stronger $\Delta \kappa$  pinning than the partially recovered microstructure of Ann3.

%Commercial Nb-Ti superconductor exhibit $J_c$ values of approx 100 $\times$ 10$^{8}$~A/m$^{2}$ in zero field and 10 $\times$ 10$^{8}$~A/m$^{2}$ in 7 T \cite{vzr}.

\begin{figure}
\includegraphics[width= 1\columnwidth]{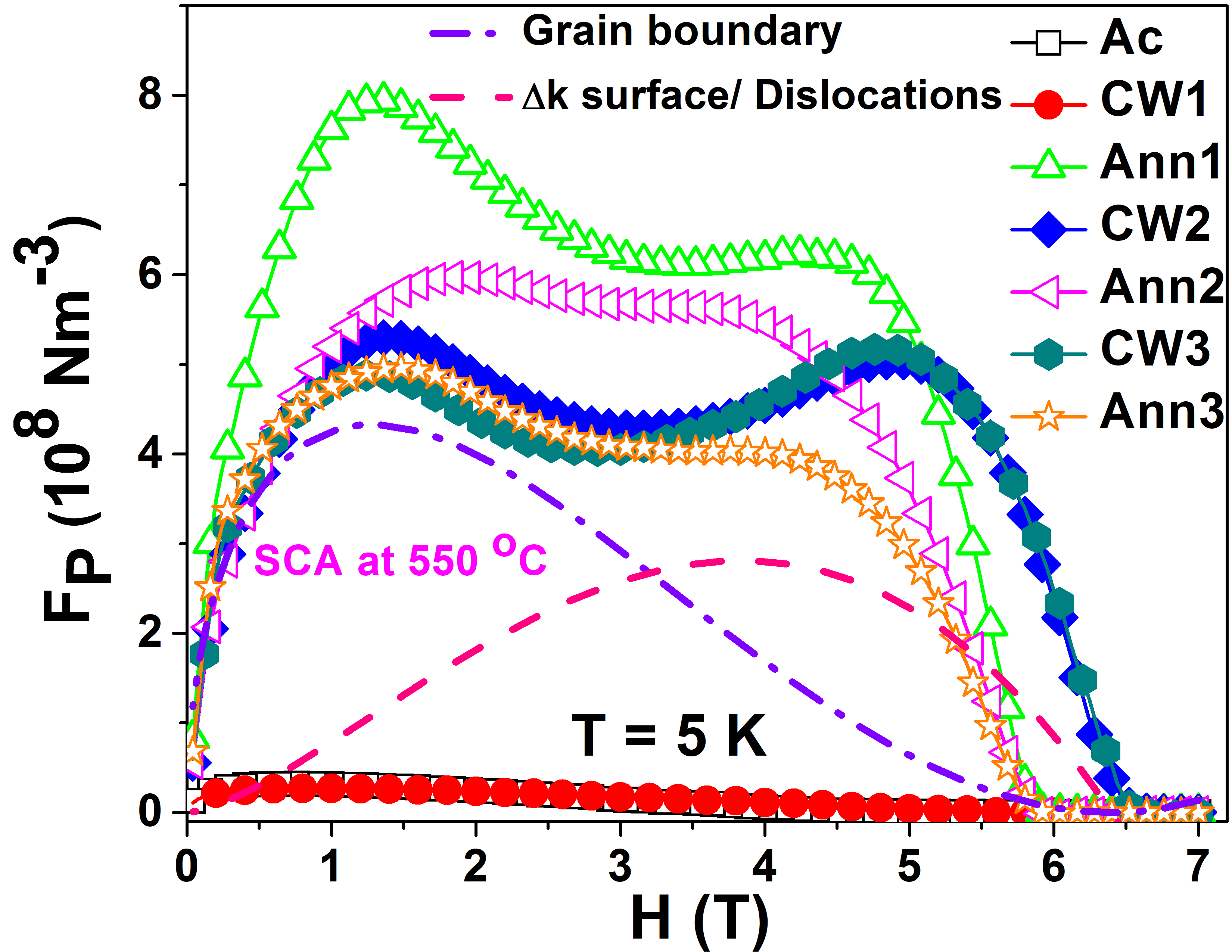}

\caption{Isothermal field dependence of the pinning force density of the V$_{0.59}$Ti$_{0.40}$Ce$_{0.01}$ alloys at different stages of SCA at 550~$^{\circ}$C, measured at 5 K. Grain-boundary pinning dominates in the low-field regime, whereas $\Delta\kappa$ pinning is dominant in high magnetic fields. The two pinning contributions shown for the Ann1 sample are representative of the behaviour observed for the other samples.}

\label{fp}

\end{figure}

The isothermal field dependence of the pinning force density ($F_P$) at 5 K for the V$_{0.59}$Ti$_{0.40}$Ce$_{0.01}$ alloy at different stages of SCA at 550~$^{\circ}$C is shown in Fig.~\ref{fp}. The measurement temperature is selected to be 5 K so that the $J_c$ can be estimated up to the magnetic irreversibility field ($H_{irr}$). $H_{irr}$ is taken as the field at which \textit{\textit{M(H)}} for increasing \textit{H} cycle bifurcates from that of the decreasing \textit{H} cycle. The magnetic field dependence of $F_P$ is analysed using the model proposed by Dew-Hudges \cite{dew}. This model correlates the $F_P$ of a superconductor with its microstructure. It proposes that $F_P \propto h^p (1-h)^q$, where $h = H/H_{\mathrm{irr}}$ is the normalized field, and the exponents $p$ and $q$ depend on the flux line pinning mechanism \cite{dew, ekin}.

The $F_P(H)$ curves in Fig.~\ref{fp} exhibit two distinct maxima, suggesting that more than one pinning mechanism contributes to the overall flux line pinning behaviour. Therefore, the $F_P(H)$ data were analysed by considering two sets of pinning mechanisms. For the Ann1 sample, the two contributions are illustrated in Fig.~\ref{fp}. $H_{\mathrm{irr}}$ for the Ann1 sample was determined to be 6.24 T and was used to normalize the applied magnetic field. The low-field contribution exhibits a maximum at $h\approx0.33$ (violet dash-dotted line), corresponding to the characteristic peak position for grain-boundary pinning. The high-field contribution peaks at $h\approx0.6$ (pink dashed line), consistent with \(\Delta\kappa\) surface pinning. $\Delta\kappa$ surface pinning arises from variations in the Ginzburg--Landau parameter ($\kappa$) at the interfaces of regions with different superconducting properties. In the present alloy, such variations occur at dislocations and $\beta$-$\alpha'$ phase boundaries \cite{sca-650}. The same two pinning mechanisms contribute to the flux line pinning behaviour of all the present samples.

\subsection{Comparative Analysis of SCA at Different Annealing Temperatures}

In the V$_{0.59}$Ti$_{0.40}$Ce$_{0.01}$ alloy, the $\beta$, $\omega$, $\alpha'$, and cerium oxide phases are present at every stage of the SCA process for all the annealing temperatures. The only difference is the appearance of the $\alpha$ phase after SCA at 550~$^\circ$C and 650~$^\circ$C, whereas it is absent after SCA at 450~$^\circ$C. Across all SCA series, the grain boundaries, dislocations and $\beta$--$\alpha'$ phase boundaries remain the dominant flux line pinning centers \cite{sca-450, sca-650}.

\begin{figure}
\includegraphics[width= 1\columnwidth]{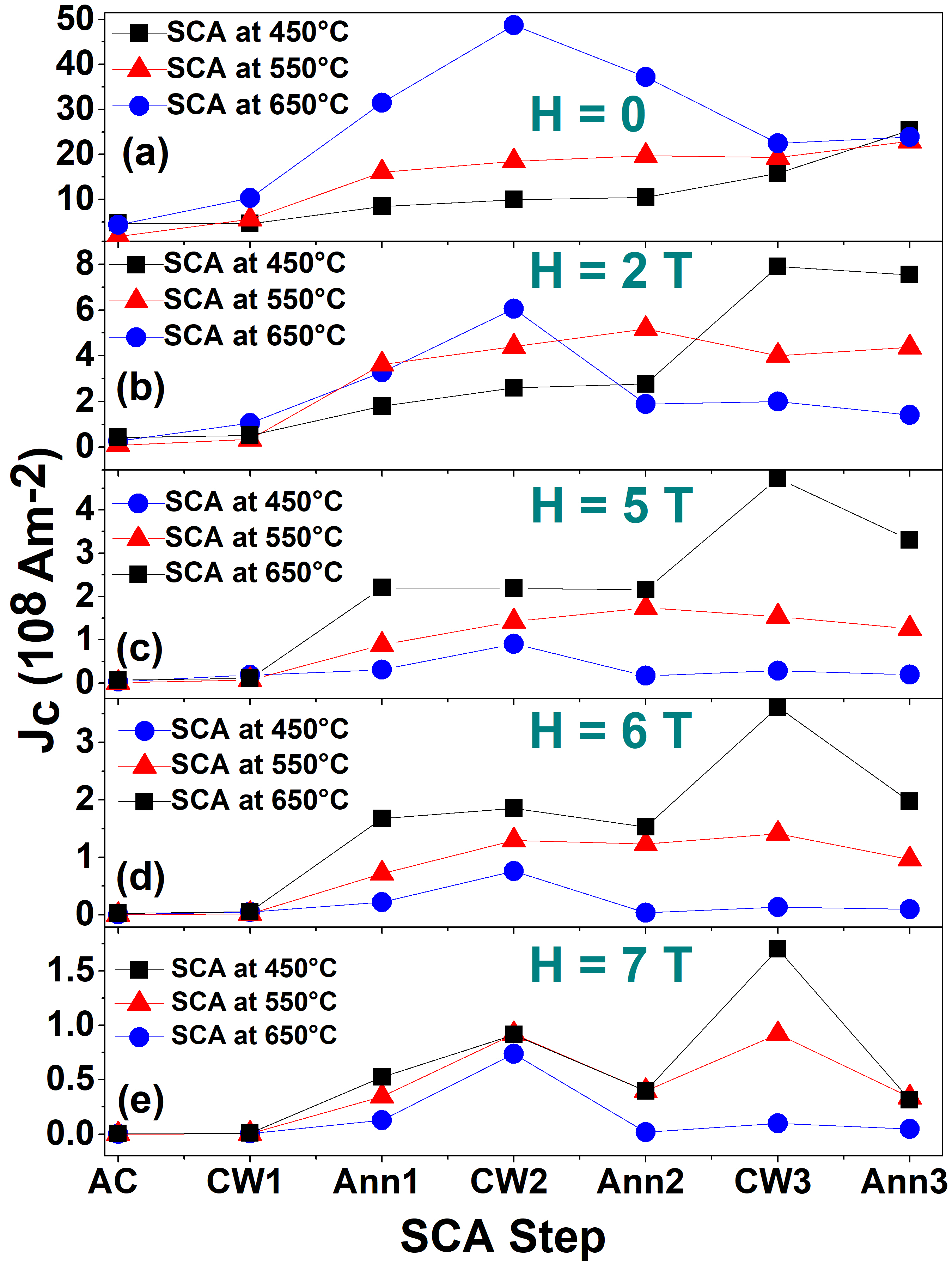}

\caption{Evolution of the $J_c$, in (a) \(H=0\), (b) \(H=2\)~T, and (c)-(e) \(H=5-7\)~T at different stages of SCA for all three SCA series. The CW2 sample processed by SCA at 650~$^\circ$C exhibits the highest $J_c$ in the low-field region, whereas the CW3 sample processed by SCA at 450~$^\circ$C shows the highest $J_c$ in higher magnetic fields.}

\label{comparison}

\end{figure}

Figure~\ref{comparison} shows the evolution of \(J_c\) at 4~K under different applied magnetic fields for the three SCA series at each stage of the SCA. In zero field [Fig.~\ref{comparison}(a)], SCA at 650~$^\circ$C shows a sharp increase in \(J_c\) during the initial steps, reaching a maximum at CW2, followed by a progressive decrease with further processing. In case of SCA at 550~$^\circ$C, $J_c$ increases significantly up to Ann1, after that it shows a moderate and relatively steady enhancement. On the other hand, SCA at 450~$^\circ$C exhibits a more gradual increase up to Ann2, followed by a pronounced enhancement after CW3, resulting in the highest ($J_c$) for Ann3 within this series. Among the three series, SCA at 650~$^\circ$C provides the highest low-field $J_c$. This enhancement can be associated with the increased contribution of grain-boundary pinning resulting from the precipitation of the $\alpha$ phase along the grain boundaries, which strengthens grain-boundary pinning \cite{sca-650}.

However, the $J_c$ of the SCA-650~$^\circ$C series decreases more rapidly with field and becomes the lowest among the three series in the high-field region [Figs.~\ref{comparison}(b)-(e)]. This deterioration is attributed to the growth of the $\alpha$-phase precipitates beyond the vortex-core size after Ann2, which reduces their effectiveness as flux-pinning centers  \cite{sca-650}. In 5~T and above [Fig.~\ref{comparison}(c)-(e)], the SCA-450~$^\circ$C series exhibits the highest $J_c$ among the three SCA series. Additionally, up to 5~T, the $J_c$ of the annealed samples is comparable to or slightly higher than that of their preceding cold-worked samples [Figs.~\ref{comparison}(b) and (c)]. However, above 5~T [Figs.~\ref{comparison}(d) and (e)], the annealed samples begin to exhibit lower $J_c$ than their preceding cold-worked steps, and this difference becomes more pronounced at 7~T. This indicates that annealing improves the low- and intermediate-field pinning characteristics, whereas cold working becomes increasingly beneficial in the high-field region. Interestingly, despite the increasing number of cold-working steps, all the three annealed samples, Ann1, Ann2, and Ann3, exhibit nearly comparable $J_c$ values at 7~T within each SCA series. 

Additionally, $J_c$ values of CW3 samples are inversely proportional to the annealing temperature. Cold working introduces a high density of dislocations, which can act as effective flux-pinning centers, particularly in the high-field region. Annealing reduces the dislocation density through recovery, and the extent of this recovery generally increases with annealing temperature. Consequently, fewer dislocations remain after annealing at higher temperatures, resulting in a lower contribution from dislocation-related pinning. Thus, although subsequent cold working reintroduces dislocations, the microstructural state from which the final cold-working step begins depends on the preceding annealing temperature, leading to a lower high-field ($J_c$) for samples subjected to higher-temperature annealing. This interpretation is consistent with the field dependence of the pinning-force density shown in Fig.~\ref{fp}, where the CW samples at 5~K show significant pinning up to approximately 6.5 T, whereas the corresponding annealed samples show a more pronounced reduction in pinning below 6~T.

\begin{figure}
\includegraphics[width= 1\columnwidth]{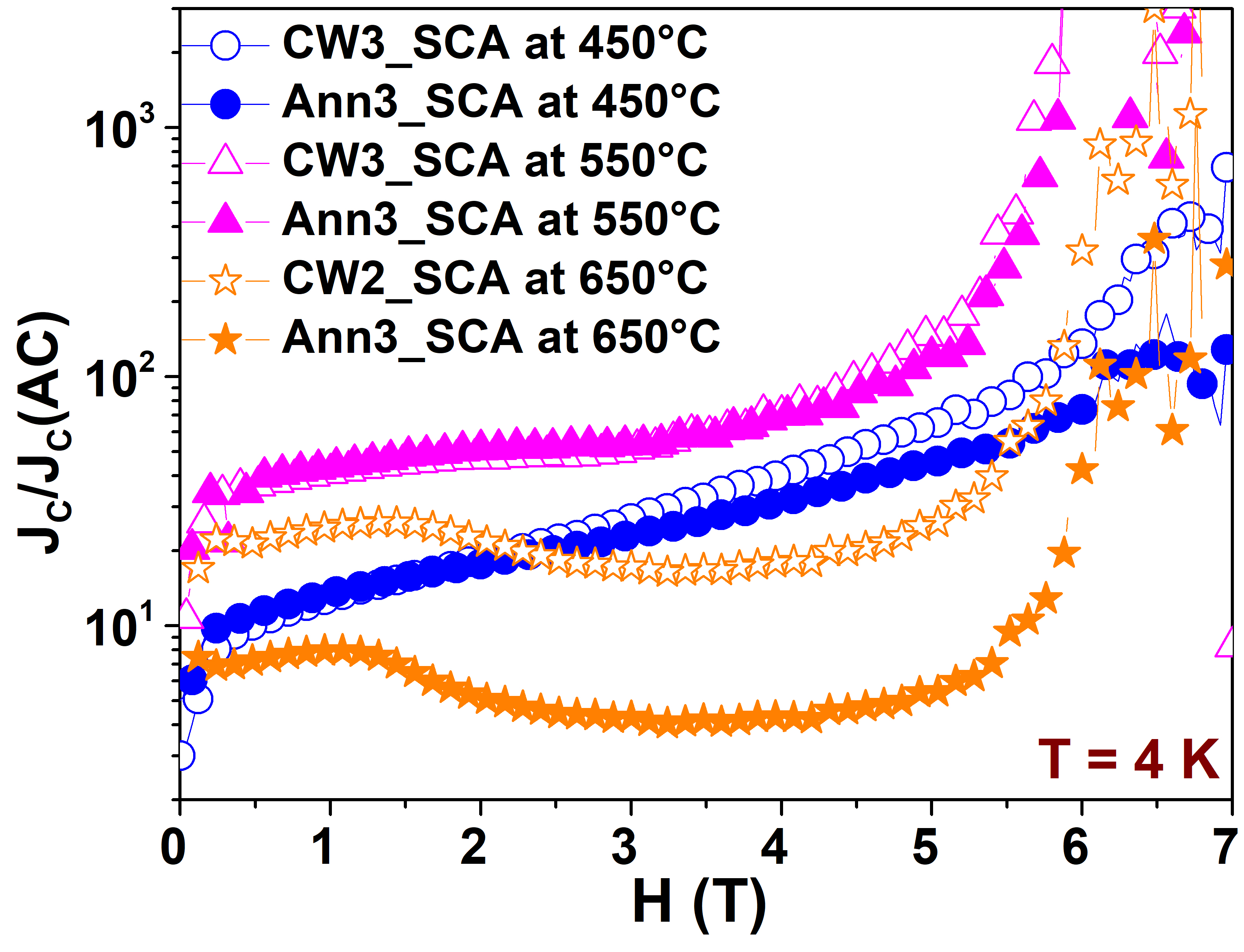}

\caption{Field dependence of the normalized critical current density at 4 K for the two best-performing samples from each SCA series. The $J_c$ values are normalized with respect to the corresponding as-cast sample of each series. SCA at 550~$^\circ$C exhibits the largest enhancement in $J_c$ over the entire measured field range.}

\label{norm}

\end{figure}

Although all alloys were prepared using the same procedure, slight variations in the cooling conditions during arc melting can lead to small differences in the microstructure of the as-cast alloys \cite{Lutjering}. Consequently, the initial $J_c$ of the as-cast sample is not identical across all series, as seen in Fig. \ref{comparison} (a). Since the enhancement produced by a particular SCA route depends on the starting state of the alloy, a comparison based solely on the absolute $J_c$ values may be misleading. Therefore, to evaluate the effectiveness of the different SCA routes, the $J_c$ values were normalized with respect to the corresponding as-cast $J_c$ of each series. Figure~\ref{norm} presents the normalized $J_c$ values for the two best-performing sample from each SCA series. SCA at 550~$^\circ$C exhibits the largest enhancement over the entire measured field range, indicating that it is the most effective processing route among those investigated.

\section{Conclusion}

The effect of SCA at 550~$^{\circ}$C on the microstructure and superconducting properties of the V$_{0.59}$Ti$_{0.40}$Ce$_{0.01}$ alloy has been investigated and compared with those reported previously for SCA at 450~$^{\circ}$C and 650~$^{\circ}$C. Annealing at or above 550~$^{\circ}$C promotes the formation of the $\alpha$-phase along the grain boundaries, thereby enhancing the grain boundary pinning. While the SCA at 650~$^\circ$C exhibits the highest $J_c$ in the low-field region, excessive growth of the $\alpha$ phase after Ann2 beyond the vortex diameter weakens the $\Delta\kappa$ pinning at the $\beta$-$\alpha$ and $\alpha$-$\alpha'$ phase boundaries, thereby reducing the high-field $J_c$. Recrystallization during annealing further reduces the dislocation density, which also contributes to the degradation of high-field $J_c$. Although the highest absolute $J_c$ is obtained after the third SCA cycle at 450~$^\circ$C, the first annealing at 550~$^\circ$C yields a comparable $J_c$ and exhibits significantly high $J_c$ values even above 7 T. Among the three SCA series, the SCA-550~$^\circ$C route produces the largest enhancement in $J_c$ relative to the as-cast alloy. The $J_c$ values in 7~T for the final cold-worked samples decrease with increasing annealing temperature, while all the annealed samples within each SCA series exhibit nearly similar $J_c$ values. The higher $J_c$ of the samples subjected to lower-temperature annealing is attributed to the higher density of dislocations retained after annealing, which act as effective pinning centers in the high-field region. Overall, annealing at 550~$^\circ$C provides an optimum balance between the formation of effective pinning defects and their preservation during annealing, resulting in an efficient flux-pinning landscape and superior overall superconducting performance of the V$_{0.59}$Ti$_{0.40}$Ce$_{0.01}$ alloy.

\section{ Data Availability Statement:} The data that support the findings of this study are available from the corresponding author upon reasonable request.

%\section{references} 

%\section{Bibliography}

%\bibliographystyle{model2-names}

%\bibliography{cas-refs}

\begin{thebibliography} {}

       \bibitem{strickland} N.M. Strickland, M. Goddard-Winchester, C. Shellard, S.C. Wimbush, J.R. Olatunji, B.E. Pavri, X. Huang, B.P.P. Mallett, K. Bouloukakis, B. Parkinson, N.J. Long, A.A. Rao, R. Pollock,
Superconductivity 17 (2026) 100231.

     \bibitem{ruiz} H.S. Ruiz, J. Hänisch, M. Polichetti, A. Galluzzi, L. Gozzelino, D. Torsello, S. Milošević-Govedarović, J. Grbović-Novaković, O.V. Dobrovolskiy, W. Lang, G. Grimaldi, A. Crisan, P. Badica, A.M. Ionescu, P. Cayado, R. Willa, B. Barbiellini, S. Eley, A. Badía--Majós, Prog. Mater. Sci. 155 (2026) 101492.

      \bibitem{garcia} L. García-Tabarés, F. Toral, J. Munilla, L. González, T. Puig, X. Obradors, Riv. Nuovo Cim. 48 (2025) 435--536.
        
        \bibitem{kingham} D. Kingham, M. Gryaznevich, Phys. Plasmas 31 (2024) 042507.


       \bibitem{prikhna} T. Prikhna, V. Sokolovsky, V. Moshchil, Materials 17 (2024) 2787.

       \bibitem{kushwaha} R.K. Kushwaha, S. Jangid, P. Mishra, S. Sharma, R.P. Singh, Mater. Adv. 7 (2026) 2379--2389.
        
          \bibitem{Banno2023} N. Banno, Superconductivity 6 (2023) 100047.      
     \bibitem{Gajda2025} D. Gajda, J. Appl. Phys. 138 (2025) 233904.

        
    \bibitem{Nishimura}  A. Nishimura, Y. Hishinuma, H. Oguro, S. Awaji, Nucl. Mater. Energy 38 (2024) 101603. 

           
        \bibitem{kitagawa} J. Kitagawa, Y. Mizuguchi, T. Nishizaki, Eur. Phys. J. B 98 (2025) 73.


        \bibitem{godeke} A. Godeke, G. Lumsden, N. Strickland, A. Otto, Supercond. Sci. Technol. 39 (2026) 045013.

        \bibitem{sharma} N. Sharma, K. Kargeti, N. Sharma, P. Chourasia, B. Vignolle, O. Toulemonde, T. Chakraborty, S.K. Panda, S. Marik, Phys. Rev. B 112 (2025) 224515.

        \bibitem{Baumann2026} J. Baumann, S.C. Hopkins, J. Dular, F. Magnus, M. Wozniak, J.D. Bijlsma, B. Medina-Clavijo, T. Boutboul, IEEE Trans. Appl. Supercond. 36 (2026) 1--9.
         
          \bibitem{jangid} S. Jangid, P.K. Meena, R.K. Kushwaha, S. Srivastava, P. Manna, S. Sharma, P. Mishra, R.P. Singh, Phys. Rev. Mater. 9 (2025) 094804.


                     
         \bibitem{aj} A. Jetybayeva, A. Mukanova, A. Nurpeissova, Z. Bakenov, V. Petrykin, S. Lee, iScience 28 (2025) 113260.
        
        
        \bibitem{guo} Z. Guo, L. Chen, Y. Li, X. Xia, G. Lin, P. Hu, D. Gong, D. Wang, Y. Ma, Materials 18 (2025) 4988.    
        
 \bibitem{sene} F. Carvalho de Castro Sene, Superconductivity 9 (2024) 100083.  



          \bibitem{idc} R. Idczak, W. Nowak, B. Rusin, R. Topolnicki, T. Ossowski, M. Babij, A. Pikul, Materials 16 (2023) 5814.

\bibitem{Lei2023} Z. Lei, C. Yao, W. Guo, D. Wang, Y. Ma, Materials 16 (2023) 1786.

 
  
        \bibitem{Nag} T. Nagasaka, T. Sugawara, S. Sakurai, K.-I. Fukumoto, Y. Yamauchi, K. Katayama, H. Watanabe, V. Tsisar, Defect Diffus. Forum 446 (2026) 21--34.

  \bibitem{butt} L. Butt, A. Dickinson-Lomas, M. Freer, J. Lim, Y.-L. Chiu, Fusion Eng. Des. 210 (2025) 114739.
        
        
            \bibitem{Zhang} Q. Zhang, L. Li, H. Huang, S. Chen, N. Jia, Y. Dong, X. Guo, K. Jin, Y. Xue, J. Nucl. Mater. 596 (2024) 155078.      
      

        
        \bibitem{tak10} T. Takeuchi, H. Takigawa, N. Banno, M. Nakagawa, M. Iwatani, K. Inoue, Y. Hishinuma, A. Nishimura, AIP Conf. Proc. 1219 (2010) 263–270. 

        \bibitem{tai} M. Tai, K. Inoue, A. Kikuchi, T. Takeuchi, T. Kiyoshi, Y. Hishinuma, IEEE Trans. Appl. Supercond. 17 (2007) 2542–2545

       \bibitem{mat15}	Md. Matin, L. S. Sharath Chandra, M. K. Chattopadhyay, R. K. Meena, R. Kaul, M. N. Singh, A. K. Sinha, S. B. Roy, Physica C 512 (2015) 32–41.
        
       		
        \bibitem{mat13} Md. Matin, L. S. Sharath Chandra, M. K. Chattopadhyay, R. K. Meena, R. Kaul, M. N. Singh, A. K. Sinha, S. B. Roy, J. Appl. Phys. 113 (2013) 163903. 

        
        \bibitem{vzr} L. S. Sharath Chandra, S. Paul, A. Khandelwal, V. Kaushik, A. Sagdeo, R. Venkatesh, K. Kumar, A. Banerjee, M. K. Chattopadhyay, J. Appl. Phys. 126 (2019) 183905, and references therin.
        
        
        \bibitem{speller} S.C. Speller,  Contemp. Phys. 66 (2025) 91--115.

                
        \bibitem{ji} X. Ji, S. Xiang, M. Zeng, S. Hu, J. Mater. Eng. Perform. 33 (2024) 854--863.
   

  \bibitem{yan} C. Yan, C. Xue, H. Han, P. Zhang, Supercond. Sci. Technol. 39 (2026) 025023.
  
\bibitem{ersoz} T.T. Ersoz, A.M.A. Mohamed, M. Jeong et al., Prog. Addit. Manuf. 11 (2026) 801--812. 
          

        
      
\bibitem{kumar} G. Kumar, M. Dahiya, R. Kumar, D. Kumar, N. Khare, Appl. Phys. A 129 (2023) 291.
        
       
 \bibitem{LPBf} T.T. Ersoz, A.E.-M.A. Mohamed, U. Mahmud, M. Jeong, Y.-L. Chiu, M.M. Attallah, Mater. Des. 258 (2025) 114681.

           
\bibitem{hidayati} R. Hidayati, J.H. Kim, G. Kim, J.H. Yun, J.-S. Rhyee, Curr. Appl. Phys. 59 (2024) 169--181.
        
        \bibitem{kim} J. Kim, S.-G. Jung, Y. Han, J.H. Kim, J.-S. Rhyee, S. Yeo, T. Park, J. Mater. Sci. Technol. 189 (2024) 60--67.
        
         
        \bibitem{shadab} M. Shadab, Y. Xing, J. Noudem, M. Miryala, J. Mater. Sci.: Mater. Electron. 35 (2024) 2236.
        
\bibitem{simon2024} G. Simon, M. Miryala, J. Alloys Compd. Commun. 3 (2024) 100023.
        
\bibitem{geng} Z. Geng, H. Oguro, J. Adv. Sci. 36 (2024) 109.




     \bibitem{yz}Y. Zhang, S. Zhang, J. Liu, F. Yang, C. Li, J. Li, P. Zhang, Chin. Phys. Lett. 41 (2024) 117402.

\bibitem{zhu} Y. Zhu, Q. Guo, P. Zhang, K. Zhang, R. Wang, Z. Zhou, L. Han, S. Wang, B. Wu, J. Li, X. Liu, Y. Feng, IEEE Trans. Appl. Supercond. 35 (2025) 6000505.

    
        
        
           

        
        
        \bibitem{gd}	S. Paul, SK. Ramjan, R. Venkatesh, L. S. Sharath Chandra, M. K. Chattopadhyay, IEEE Trans. Appl. Supercond. 31 (2021) 8000104.
        
        
        \bibitem{y}	SK. Ramjan, L. S. Sharath Chandra, R. Singh, P. Ganesh, A. Sagdeo, M. K. Chattopadhyay, J. Appl. Phys. 131 (2022) 063901. 


        \bibitem{si} A. Khandelwal, L. S. Sharath Chandra, A. Sagdeo, R. Singh, M. Gangrade, R. Venkatesh, M. K. Chattopadhyay, Mater. Sci. Eng. B 317 (2025) 118158.

        
        \bibitem{sri} N. Srivastava, G.A.B. Matthews, J. Liu, S.C. Speller, C.R.M. Grovenor, S. Santra, J. Alloys Compd. 1002 (2024) 175526.
        
        \bibitem{sca-450} SK. Ramjan, A. Khandelwal, S. Paul, L. S. Sharath Chandra, R. Singh, K. Kumar, R. Rawat, S. Dutt, A. Sagdeo, P. Ganesh, M. K. Chattoadhay, J. Alloy. Compound.  976 (2024) 173321, and references therein.
        
        \bibitem{sca-650} A. Khandelwal, N. Mirza, L.S. Sharath Chandra, R. Singh, A. Sagdeo, M.K. Chattopadhyay, J. Supercond. Novel Magn. 38 (2025) 34.
        
      
        
        
        
        
\bibitem{POWDER} G. Nolze, in: Powder Diffraction: Proc. of the II Int. School on Powder Diffraction (Kolkata, India, 2002) pp. 146--155.


\bibitem{chiang} H.-W. Chiang, R.N. Blumenthal, R.A. Fournelle, Solid State Ionics 66 (1993) 85--95.
\bibitem{ceo} E. A. Kümmerle, and G. Heger, J. Solid State Chem. 147 (1999) 485-500.
               
        \bibitem{bean64} C. P. Bean, Rev. Mod. Phys. 36 (1964) 31.
        
        \bibitem{dew} D. Dew-Hughes, Philos. Mag. 30 (1974) 293.
        
        \bibitem{ekin} J. W. Ekin,  Supercond. Sci. Technol. 23 (2010) 083001.

        
\bibitem{Lutjering} G. Lütjering, J.C. Williams, A. Gysler, Microstructure and mechanical properties of titanium alloys, in: \textit{Microstructure and Properties of Materials}, vol. 2, Wiley-VCH, 2000, pp. 1--77.
        
        
        
        
        
        













    

%
%        \bibitem{fra77} G. W. Franti, D. A. Koss, On the equalibrium silicide in beta Ti-V alloys containing Si, Metall. Trans. A 8 (1977) 1639-1641.


\end{thebibliography}

%\section*{References}


\end{document}